\documentstyle[prl,aps,epsf]{revtex}

\begin{document}
\title{Cascades of energy and helicity in the GOY shell model of turbulence}
\author{P. D. Ditlevsen\\
The Niels Bohr Institute, Department for Geophysics,\\
University of Copenhagen, Juliane Maries Vej 30,\\
 DK-2100 Copenhagen O, Denmark.}
\date{July 22, 1996}
\maketitle
\begin{abstract}
The effect of extreme hyperviscous damping, $\nu k_n^p, p=\infty$
is studied numerically in the GOY shell model of turbulence.
It has resently been demonstrated [Leveque and She, Phys. Rev. Lett, 75,2690 (1995)] that the inertial range scaling in the
GOY model is non-universal and depending on the 
viscous damping. The present study shows that the deviation from
Kolmogorov scaling is due to the cascade of the second 
inviscid invariant. This invariant is non-positive definite and in this 
sense analogous to the helicity of 3D turbulent flow. 
\end{abstract}

\vspace{0.5cm}

The invicid invariants, like the energy, of the Navier-Stokes equation
are important quantities determining the dynamics of turbulent flow. The
major difference between 2D and 3D flow originates from
existence of a second inviscid invariant, enstrophy, in 2D -
absent in 3D flow. This gives strong constraints on the 2D
flow dynamics. Conservation of enstrophy leads to forward
cascade of enstrophy and backward cascade of energy in 2D,
giving rise to large-scale coherent structures in the flow. In 3D
there also exists a second quadratic inviscid invariant, namely the
helicity, defined as the integral of the scalar product of the velocity
and the vorticity.  It has been proposed to be important for 3D
turbulence\cite{Lesieur}. It was shown by Kraichnan \cite {K} 
that the interaction
between waves of opposite helicity is stronger than interactions
between waves of same helicity. Consequently, helical flow might slow down the
cascade of energy to the dissipation range\cite{Kerr}.

Since the spectral helicity density grows with a factor
$k$, the modulus of the wave-vector, over the the energy density the
helicity dissipation would, in the $R \rightarrow \infty$ limit,
prevent energy dissipation.  The situation would resemble that of 2D
turbulence where cascade of enstrophy prevents forward cascade of
energy. In 2D
turbulence the conservation of both $\int dk u(k)^2$ and $\int dk k^2
u(k)^2$ implies that forward cascade of enstrophy must be accompanied
by a backward cascade of energy. This strong constraint does, however,
not apply in the case of helicity transfer since helicity is not
positive definite. So we only have the weaker assessment mentioned
above that the efficiency of energy dissipation could be depending on
the non-linear helicity transfer and the helicity dissipation.

Considerable interest has lately been given to the GOY 
model of turbulence, introduced by
Gletzer and examined by Yamada and Ohkitani\cite{GOY}. 
Comprehensive lists of resent references on the GOY model can be found
in refs. \cite{benzi,biferale}. The behavior of the helicity in 
different shell models have resently been investigated \cite{biferale,b+k}.

The GOY model is a simplified reduced wave-number 
analog to the spectral Navier-Stokes
equation. The spectral domain is represented as shells,
each of which is defined by a wavenumber $k_n = k_0 \lambda^n$, where
$\lambda$ is a scaling parameter defining the shell spacing.
There are $2 N$
degrees of freedom, where $N$ is the number of shells, namely the
generalized complex shell velocities, $u_n$ for $n=1,N$.
The dynamical equation for the shell velocities is,

\begin{equation}
\dot{u}_n=i k_n (a u_{n+2}^*u_{n+1}^*+\frac{b}{\lambda}u_{n+1}^*u_{n-1}^*
+\frac{c}{\lambda^2}u_{n-1}^*u_{n-2}^*) -\nu k_n^p u_n + f \delta_{n,n_0},
\label{dyn}
\end{equation}

where the first term represents the non-linear wave interaction or
advection, the second term is the dissipation, and the third term the
forcing, where $n_0$ is some small wavenumber.  Throughout this paper
the standard 3D GOY model parameter values, $\lambda=2, k_0= \lambda^{-4}, 
a=1, b=c=-1/2$ is used.
The GOY model in this form contains no information about
phases between waves, thus there cannot be assigned a flow field in
real space. 

The model has two conserved integrals, in the case of
no forcing and no dissipation $(\nu = f = 0)$ (inviscid invariants). 
These are,
$E=\frac{1}{2}\sum_{n=1}^N |u_n|^2$
and
$H=\frac{1}{2}\sum_{n=1}^N (-1)^n k_n |u_n|^2$
which corresponds to the conservation of energy and a second
non-positive definite quantity interpreted as analogous to 
helicity\cite{benzi} for
the Navier-Stokes equation of 3D turbulence, hereafter referred to as
the helicity. It should be stressed that this analogy is only in
the sense that both quantities are non-positive definite. 
In this model each shell
is maximally helical with alternating sign, 
since the numerical value of the helicity density
is $k$ times the energy density.

The forcing in (\ref{dyn}) is applied at a small wave number and the
dissipation dominates at large wave-numbers, so we can define an
inertial range where the non-linear energy cascading terms dominate and
the Kolmogorov scaling arguments apply.
For an energy cascade we have the "Kolmogorov scaling" of the shell-velocities,
$|u| \sim \eta^{1/3}k^{-1/3}$,
and for a helicity cascade we have the "helicity scaling",
$|u| \sim \tilde{\eta}^{1/3}k^{-2/3}$,
where $\eta$ and $\tilde{\eta}$ are the mean dissipation per unit time of energy and 
helicity respectively.  The
model has both the Kolmogorov scaling, $u_n=k_n^{-1/3}g(n)$ and the
helicity scaling, $u_n=(-1)^nk_n^{-2/3}g(n)$ as unstable fixed points in the
unforced and inviscid case.
The function, g(n)=g(n+3), is any mod(3) function.  The mod(3) symmetry
is an artifact of the GOY model discussed in detail in ref. \cite{benzi}.  
It will become important in the following.
The Kolmogorov fixed point plays an important role for the behavior of
the GOY model, with forcing and dissipation, in the sence that the phase space
trajectory of the shell velocities seems to "curl around" this point with
the average values of the velocities close to the fixed point values.
It was shown in a numerical study by Leveque and She \cite{sl} that the 
inertial range scaling in the GOY model is not universal and depending 
on the form of the viscous damping. They attributed this non-universality
to the reflection of energy flow from the viscous subrange back into the
inertial range in the case of hyperviscosity $(p>2)$. By studying the
extreme case $p=\infty$, I suggest a different explanation for the 
non-universality of the inertial range scaling, namely that cascade of
helicity blocks the cascade of energy and thus changes the scaling. 
This effect is probably specific to the GOY model, and a consequence
of the specific odd-even asymmetry of the helicity in the GOY model.

The numerical study is performed on 2 versions of the GOY model. Firstly,
the usual model with normal $(p=2)$ dissipation and secondly, with 
viscosity only applied on outermost shell, corresponding to $p=\infty$ 
hyper-viscosity.
The forcing in (\ref{dyn}) is in both cases  
taken to be $f=f_0/u_{n_0}^*$ where $f_0$ is a constant. This gives a constant input
of energy (and helicity) per unit time.  

Figure 1 shows the result of a numerical calculation for
$f_0=(1+i)\times 10^{-3}, n_0=4$ and  $\nu =10^{-5}$.  The model
has 20 shells and is run for $3000$ time units after a $1000$ time units
spinup starting from the (unstable) Kolmogorov fixed point. First panel shows
$\log_{\lambda}(\langle |u_n|\rangle )$ as a function of the shell
number $n=log_{\lambda} (k_n)$, where $\langle .\rangle$ denotes
temporal average.  The spectral slope is close to the Kolmogorov
scaling, shown by the line, thus the model shows an energy cascade.

\begin{figure}[htb]
\epsfxsize=10cm
\epsffile{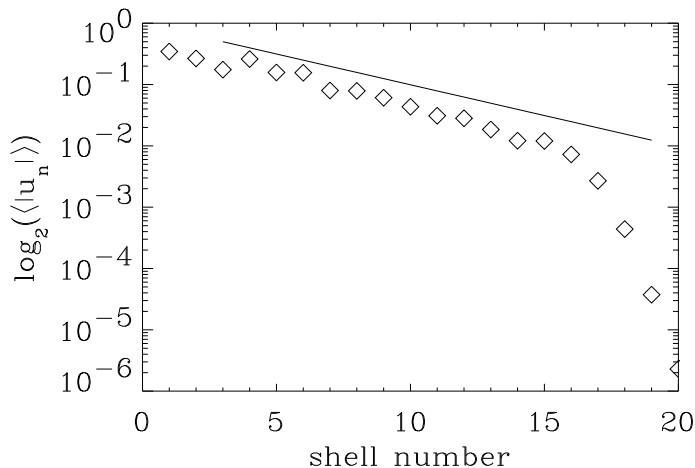}
\caption[]{ 
\label{1}
The standard GOY model with
$f_0=(1+i)\times 10^{-3}, n_0=4$ and  $\nu =10^{-5}$.  The model
has 20 shells and is as before run for $3000$ time 
units after a $1000$ time 
units
spinup starting from the (unstable) Kolmogorov 
fixed point. 
$\log_{\lambda}(\langle |u_n|\rangle )$ is shown as a 
function of the shell
number $n=log_{\lambda} (k_n)$. The line indicates 
the Kolmogorov scaling
with spectral slope of $-1/3$. 
}
\end{figure}

The sources and sinks for energy and helicity are

\begin{equation}
\dot{E}=\sum_n(u_n^*f_n-\nu k_n^2 u_n^2) 
\label{ee}
\end{equation}
and
\begin{equation}
\dot{H}=\sum_n((-1)^n
k_n u_n^*f_n-(-1)^n \nu k_n^3 u_n^2). 
\end{equation}

This means that helicity of opposite
signs is dissipated at every second shell and therefore
total helicity is produced for
odd numbered shells in the dissipation range. 

The inertial range flow is dominated by the non-linear transfer
of the conserved quantities. The key assumption by Kolmogorov (1941) is that the
inertial range flow does not depend on the specific form of the
small scale viscous dissipation. This is not the case for the 
GOY model.
In order
illustrate this point a study of a slightly modified GOY model is
presented here. The dissipation is taken to be active only on 
the outermost shell,
number $N=20$, this corresponds to hyperviscosity in 
the limit, $p\Rightarrow \infty$, with
$\nu = \nu_0 k_N^{-p}$ and $\nu_0$ held constant.  
With this choice of dissipation the behavior of the GOY model
changes dramatically even though the non-linear terms in the 
governing equation remains unchanged.

Figure 2 shows the result of a
numerical integration with the modified GOY model 
where $f_0=(1+i)\times 10^{-8}$ and $\nu_0 = 10^{-7}k_{20}^2 = 429.5$.
First panel shows the velocity spectrum where
the mod(3) symmetry of the shell model becomes
dominant. The scaling behavior of shells numbered 10,13,16,19 is different
from the rest. The line in the figure has a slope of $-2/3$ corresponding 
to the cascade of helicity rather than energy. 

However, if $U_n=|u(n-1)u(n)u(n+1)|^{1/3}$ is considered the mod(3) symmetry is
eliminated and the inertial range scaling approximately
reemerges as an arithmetic mean of the scaling behaviors for the shells
$3n, 3n+1, 3n+2$ (figure 2, second panel). The line in the figure
has a slope of $-1/2$ corresponding to equipartitioning of helicity.
The hyper-viscosity of the model, only pulling out helicity through the
(18,19,20) triade, cannot maintain a helicity cascade.

\begin{figure}[htb]
\epsfxsize=10cm
\epsffile{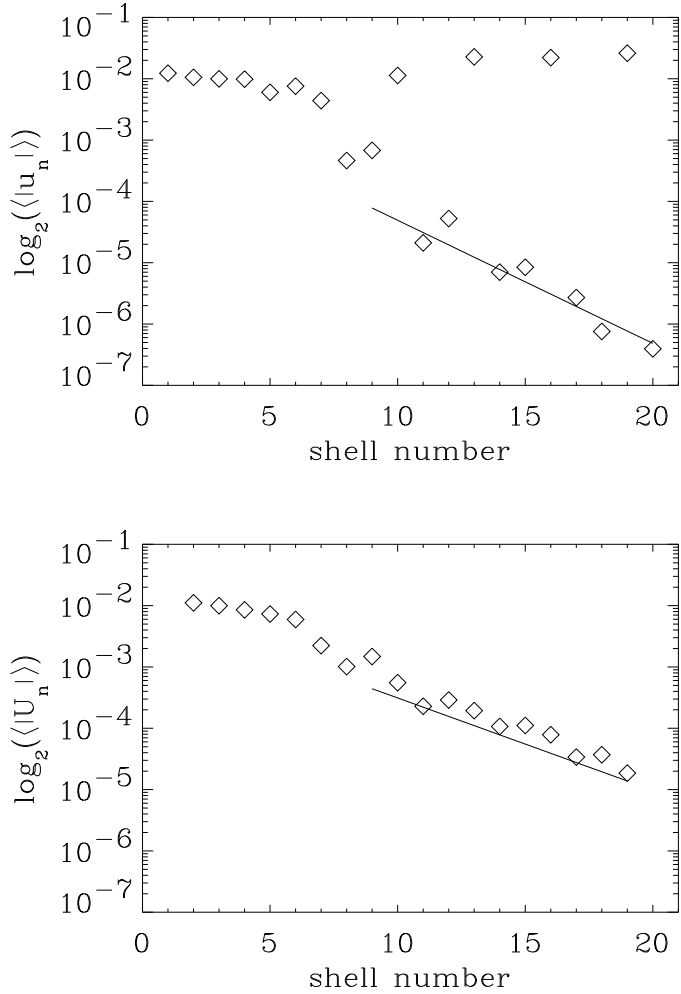}
\caption[]{ 
\label{2}
The modified GOY model 
with $f_0=(1+i)\times 10^{-8}$ and $\nu_0 = 10^{-7}k_{20}^2 = 429.5$.
This model has likewise 20 shells and is run for $3000$ time units.
First panel is the same as in Figure 1. 
The line in the figure has a slope of $-2/3$ corresponding 
to the cascade of helicity. 
Second panel shows the spectrum of $U_n=|u(n-1)u(n)u(n+1)|^{1/3}$.
The line has a slope of $-1/2$ corresponding to equipartition of 
helicity.
}
\end{figure}

The result can be interpreted as a spectral
bump at the end of the spectrum with an approximate equipartition 
of helicity. 
Similar results have been found by Borue and 
Orszag \cite{BO} for numerical simulations of the 3D
Navier-Stokes equation. In this case the accumulation seems to be an 
effect of energy not being able to be transferred across the ultra-violet
cutoff.

In the case  $(f=\nu=0)$ the model has, discarding the boundary effects, 
besides the two scaling fixed points a 
periodic solution.
It is easy to verify that the following satisfies the
dynamical equation (1):

\begin{eqnarray}
u_{3n-2} (t)= s k_0^{-1}\lambda^{-3n} \nonumber \\
u_{3n-1} (t)= \lambda^{3n\gamma}\sqrt{-\alpha_1}
e^{-i\sqrt{-\alpha_1\alpha_2}t}\nonumber \\
\label{cycle}
u_{3n} (t)= \lambda^{3n\gamma}\sqrt{\alpha_2}
e^{i\sqrt{-\alpha_1\alpha_2}t} \nonumber\\
\end{eqnarray}

with

\begin{eqnarray}
\alpha_1=s\lambda^{-4}(1+b\lambda^2+c\lambda^{-3\gamma+1}) \nonumber \\
\alpha_2=s\lambda^{3\gamma-3}(1+b\lambda^{-3\gamma-1}+
c\lambda^{-3\gamma+1}),\nonumber \\
\end{eqnarray}

and $s$ is an arbitrary constant. The scaling 
parameter, $\gamma$, is related to the scaling fixed point by
$\gamma = -(\alpha+1)/3$, where $z=\lambda^{\alpha}$ is a solution to
$1+bz+cz^2=0$, thus a generator of one of the conserved quantities\cite{ditlev}.
So there are two values possible for $\gamma$; $\gamma=
-(2+i\pi /\log(\lambda))/3$ corresponding to the generator, $z=-2$,  of helicity, or
the fluxless fixed point of the GOY model, and $\gamma=-1/3$ corresponding
to the generator, $z=1$, of energy, or the Kolmogorov fixed point.

There is in this periodic solution a complete phase locking of all the shells.
The energy and helicity fluxes are both zero in the periodic solution,
so in some sence it corresponds to the fluxless
fixed point of the GOY model.

The shell-velocities of shells 8, 11, 14, 17, 20 are out of phase with 
those of shells 9, 12, 15, 18, while shell-velocities of shells 10, 13, 16, 19 are almost constant.
From this it seems as if the periodic solution plays the same role for this 
modified GOY model as does the Kolmogorov fixed point for the GOY model.
This could indicate the existence of an (unstable) limit cycle in the 
modified GOY model.

The dissipation of helicity is irregular, associated with the bursts
in shell number 20 (and all other shells), the energy dissipation is completely 
blocked, resulting in a steady increase of energy concentrated
on shells 10,13,16,19. The mechanism for preventing the forward cascade
of energy in this model is analogous to the 2D case, where the forward cascade
of enstrophy governs the dynamics.

The system shows no signs of approaching a 
statistical equilibrium simply because the helicity can grow to
arbitrarily large negative values (dominated by shell number 19).
The reason for this can be understood by examining the
dissipative energy balance.
In the case of normal viscosity it follows from (\ref{ee}) that
the phase space trajectory will 
be attracted to the hyper-ellipsoid given by $\sum \nu k_n^2 |u_n|^2 = f$
which is a compact $2N-1$ dimensional object. In the $p = \infty$
case this object will be the $2N-1$ dimensional "hyper-cylinder" defined by
$|u_{20}|= f/\nu k_{20}^2)$. This is not compact and ergodicity
does not apply, thus the $p\rightarrow \infty$ is a singular limit and
no statistical equilibrium can be reached.
The relative change in energy over the integration is of the order $10^{-2}$
such that the system is in a quasi-equilibrium state where 
 statistical equilibrium is reestablished in the limit where the
energy injection rate goes to zero. Taking an even smaller forcing does not change the 
statistics, thus the statistical timeaveraging is meaningful.

If the energy cascade is effectively blocked the energy should be 
pulled out of the system by a drag at the small wave numbers corresponding
to to backward cascade in 2D. This is not seen, so an invers energy cascade cannot
be established. The model was also run with the usual dissipation 
but only active on every second shell. The result of this run was 
essentially the same as for the modified GOY model.

In conclusion we see that the non-linear transfer in the GOY
model depends crucially on the dissipation properties for
both conserved quantities, energy and helicity. 
It is still an open question how much the GOY model reflects
the dynamics of the Navier-Stokes equation. These findings 
certainly indicates that the model is to restricted in some
sence to represent real flow. 
The dissipative term introduced in the modified model is
certainly not very realistic, in real flow
helicity of both signs will be dissipated, which it is not
in the modified model presented here.
The findings from this simple model
might indicate that in the inertial range flow
the conservation of helicity could block the energy cascade and
thus alter the Kolmogorov $k^{-5/3}$ scaling.  

Acknowledgements: I would like to thank A. Wiin-Nielsen, J. R. Herring
and R. M. Kerr for valuable discussions. The work was granted
by the Carlsberg foundation.


\end{document}